\documentclass[
  aip,
  adv,
  superscriptaddress,
  amsmath,
  amssymb,
  reprint,
  raggedbottom,
  article-title,
  longbibliography
]{revtex4-2}

\usepackage{graphicx}
\usepackage{dcolumn}
\usepackage{bm}
\usepackage[utf8]{inputenc}
\usepackage[T1]{fontenc}
\usepackage{mathptmx}
\usepackage{etoolbox}
\usepackage{comment}

\graphicspath{{figures/}}

\newcommand{\missingplot}[1]{%
  \fbox{\parbox[c][3cm][c]{0.9\linewidth}{\centering\footnotesize
  Missing figure:\par\texttt{\detokenize{#1}}}}}
\newcommand{\plotfile}[1]{%
  \IfFileExists{figures/#1}{\includegraphics[width=\linewidth]{figures/#1}}{%
    \IfFileExists{#1}{\includegraphics[width=\linewidth]{#1}}{\missingplot{#1}}}}
\newcommand{\panel}[3]{%
  \begin{minipage}[t]{#1\linewidth}
    \vspace{0pt}\raggedright\textbf{(#2)}\par\smallskip
    \centering\plotfile{#3}
  \end{minipage}}
\newcommand{\centeredpanel}[3]{%
  \begin{minipage}[c]{#1\linewidth}
    \raggedright\textbf{(#2)}\par\smallskip
    \centering\plotfile{#3}
  \end{minipage}}

\begin{document}
\title{Magnetic Field Sensing with Actuated Stochastic Magnetic Tunnel Junctions}
\author{Sohum Udani\textsuperscript{*}}
\email{sau2090@nyu.edu}
\affiliation{Center for Quantum Phenomena, Department of Physics, New York University, New York, New York 10003, USA}
\author{Troy Criss}
\affiliation{Center for Quantum Phenomena, Department of Physics, New York University, New York, New York 10003, USA}
\author{Andrew Haas}
\affiliation{Department of Physics, New York University, New York, New York 10003, USA}
\author{Daniel Stein}
\author{Andrew D. Kent}
\affiliation{Center for Quantum Phenomena, Department of Physics, New York University, New York, New York 10003, USA}

\begin{abstract}
Stochastic sensors encode physical signals in the probabilities of random outcomes, enabling sensing through repeated measurements of discrete states. Here, we demonstrate magnetic field sensing with electrically actuated stochastic magnetic tunnel junctions, in which a write pulse generates a stochastic binary outcome. Each measurement cycle prepares a perpendicular magnetic tunnel junction (pMTJ) in a known magnetic state, applies a write pulse, and electrically reads the resulting state. An applied field changes the state occupation probability, allowing both static and time-varying fields to be inferred from binary outcomes. Static fields are detected through shifts in the write voltage required to achieve a 50\% parallel-state occupation probability, while AC fields produce probability oscillations at fixed pulse settings. We detect AC fields from 5 to 100~Hz in a 50-nm-diameter pMTJ, with a minimum field-equivalent noise amplitude spectral density of approximately \(9~\mu\mathrm{T}/\sqrt{\mathrm{Hz}}\) at 100~Hz for a 10.6~MHz measurement-cycle rate. This is about twice the \(4.6~\mu\mathrm{T}/\sqrt{\mathrm{Hz}}\) expected from binomial statistics for statistically independent measurement cycles. We discuss routes to improve sensitivity by increasing the measurement-cycle rate, which lowers the binomial noise floor, and by increasing the field dependence of the state occupation probability. We also identify resonant activation as a direction for future investigation.
\end{abstract}
\maketitle

\section{Introduction}

Magnetic field sensors with small active volumes enable local probing of magnetic materials and high-resolution imaging of magnetic fields, as in scanning SQUID and scanning Hall probe microscopy~\cite{Kirtley1995,Kent1994,Sandhu2004Micro,Sandhu2004Nano}. A variety of sensing technologies have been developed, including superconducting quantum interference devices (SQUIDs), atomic magnetometers, Hall sensors, and nitrogen-vacancy (NV) centers in diamond~\cite{Popovic2004,Bennett2021}. These approaches differ in their sensitivity, sensing volume, operating conditions, and readout requirements, and comparisons across technologies show a trade-off between field sensitivity and active sensing volume~\cite{Forstner2012PRL}. The most sensitive magnetometers require cryogenic cooling, as for SQUIDs~\cite{Kirtley1995,Faley2006,Baudenbacher2003}, or optically pumped atomic vapors, as for spin-exchange-relaxation-free (SERF) magnetometers~\cite{Dang2010,Maser2011}. NV centers reach nanoscale sensing volumes but rely on optical excitation and readout~\cite{Maze2008,Balasubramanian2009}, while Hall sensors provide room-temperature electrical readout and have been scaled to dimensions of 50~nm~\cite{Sandhu2004Nano}. These considerations motivate nanoscale sensors in which a magnetic element provides the field response and both control and readout are electrical.

Magnetic tunnel junctions (MTJs) offer a platform for such measurements because their electrical resistance depends on the relative magnetization directions of two magnetic layers separated by an insulating barrier. In conventional MTJ sensors, an applied field rotates the free-layer magnetization, and the resulting resistance change is read continuously~\cite{Davies2021}. The noise of such sensors includes low-frequency \(1/f\) and thermal magnetic contributions that grow as the junction is made smaller~\cite{Egelhoff2009}, which limits analog readout in nanoscale devices. Field information can also be obtained from the statistics of transitions between magnetic states, an approach well suited to nanoscale junctions, in which thermal fluctuations are significant.

A stochastic magnetic field sensor encodes the applied field in the probability of its binary output. Although an individual outcome does not uniquely determine the field, repeated measurements reveal the field-dependent output probability, allowing a continuous-valued field to be inferred from discrete states. Stochastic MTJs fall into two classes~\cite{Sun2026}. In superparamagnetic tunnel junctions, the energy barrier is low enough that thermal fluctuations drive spontaneous transitions, and the field can be inferred from state probabilities and dwell times~\cite{Hirama2025,Liu2025}. Their performance then depends on the transition timescales, because correlations between successive observations limit the number of statistically independent measurements. In actuated, or stochastic-write, MTJs~\cite{Fukushima2014,Rehm2023}, the barrier suppresses spontaneous reversal. Each measurement cycle resets the junction to a known state, applies a write pulse that switches it with a tunable probability, and reads the result. Successive outcomes are then expected to be nearly statistically independent. The outcome rate is set by the pulse sequence, and the write pulse amplitude and duration set the operating point electrically. Field-programmable gate array (FPGA)-controlled platforms have used this pulse-induced stochastic switching to generate random bits~\cite{Dubovskiy2024,Criss2026}, at rates up to 104~MHz~\cite{Sidi2025}. Because the switching probability depends on the applied field, the same platforms can also serve as magnetic field sensors.

\begin{figure*}[t]
  \centering

  \begin{minipage}[c]{0.68\textwidth}
    \centering

    {\raggedright\textbf{(a) Experimental Setup}\par}
    \vspace{0.2em}
    \includegraphics[width=\linewidth]
      {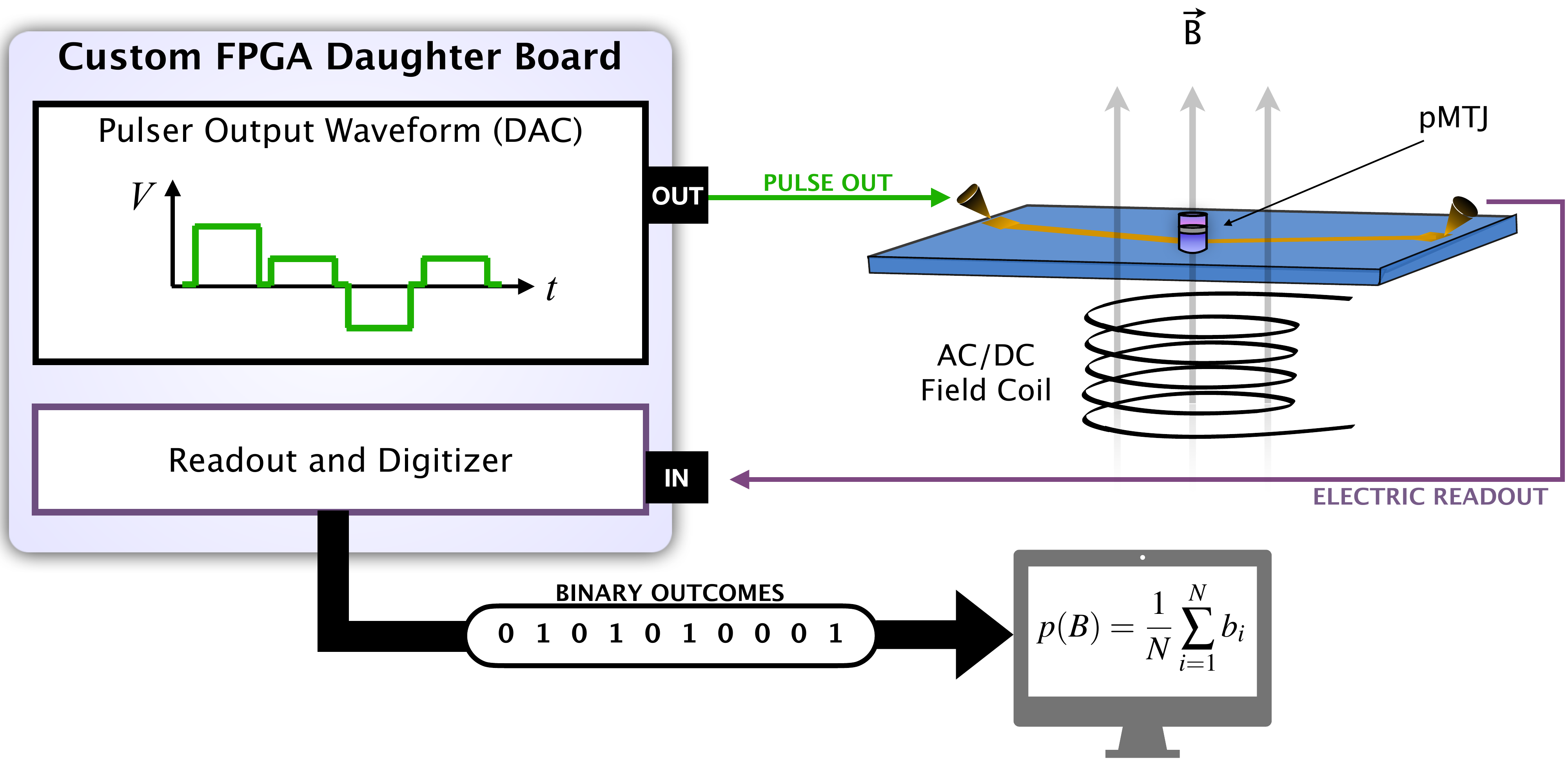}

    \par\vspace{1.2em}

    {\raggedright\textbf{(b) Device States and Measurement Cycle}\par}
    \vspace{0.2em}
    \includegraphics[width=\linewidth]
      {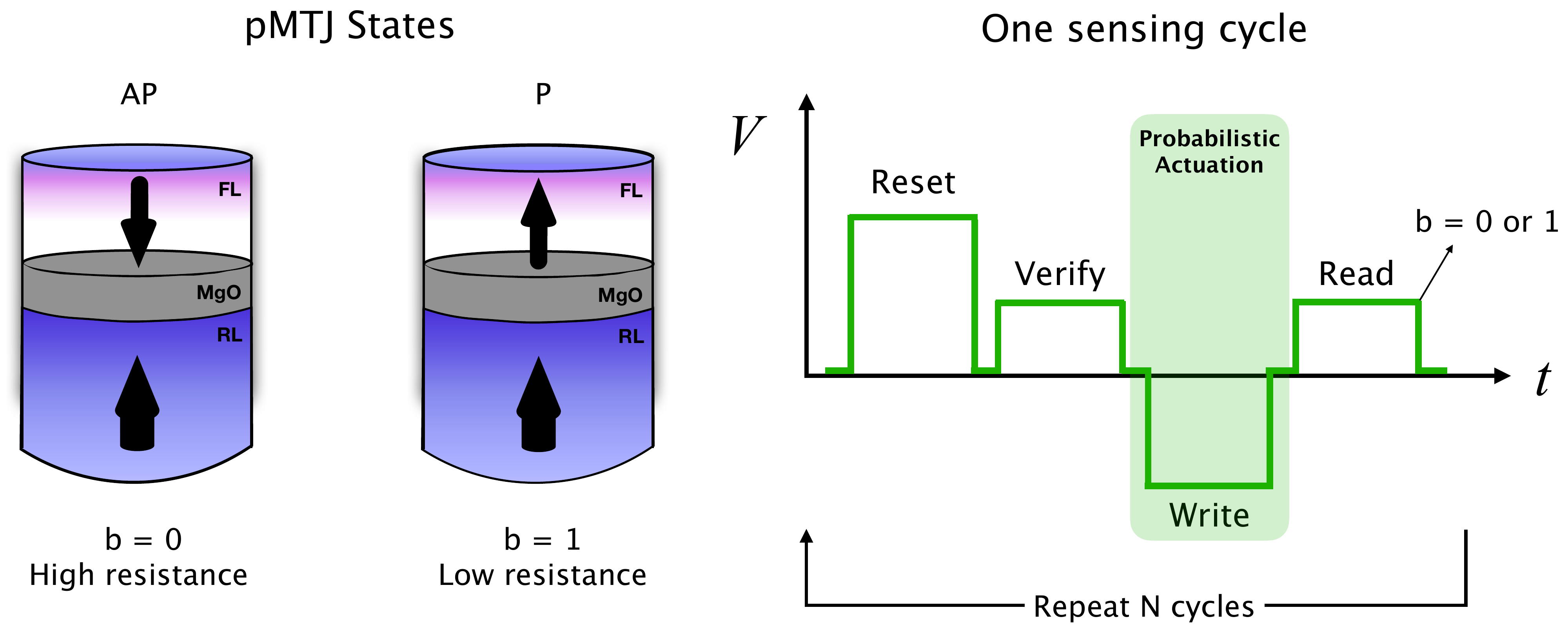}
  \end{minipage}%
  \hspace{0.03\textwidth}%
  \begin{minipage}[c]{0.29\textwidth}
    \setlength{\abovecaptionskip}{0pt}
    \setlength{\belowcaptionskip}{0pt}

    \caption{Experimental setup and measurement protocol for magnetic field sensing with actuated stochastic pMTJs.
    \textbf{(a)} The FPGA and custom analog daughter board generate electrical pulses and read the pMTJ resistance state. An electromagnet beneath the chip applies DC or AC magnetic fields perpendicular to the film plane. Binary outcomes are transferred to a computer and averaged to estimate the P-state occupation probability.
    \textbf{(b)} Antiparallel (AP) and parallel (P) configurations of the free layer (FL) and reference layer (RL), separated by the MgO tunnel barrier. We encode the high-resistance AP state as $b=0$ and the low-resistance P state as $b=1$. Each measurement cycle consists of \textit{reset}, \textit{verify}, \textit{write}, and \textit{read} pulses. The \textit{write} pulse produces a stochastic magnetic-state outcome, which is determined by the subsequent read pulse.}
    \label{fig:setup}
  \end{minipage}

\end{figure*}

In this work, we use electrically actuated stochastic perpendicular magnetic tunnel junctions (pMTJs) as magnetic field sensors. We detect static fields through shifts in the write voltage required for a 50\% probability of occupying the parallel (P) state, and AC fields from 5 to 100~Hz through probability oscillations at fixed pulse settings. In a 50-nm-diameter pMTJ with a free-layer volume of approximately \(3.9\times10^{-24}~\mathrm{m}^3\), the minimum field-equivalent noise amplitude spectral density is approximately \(9~\mu\mathrm{T}/\sqrt{\mathrm{Hz}}\) at 100~Hz. This is about twice the binomial benchmark for statistically independent measurement cycles at our 10.6~MHz measurement-cycle rate. Our analysis shows how the field response of the P-state occupation probability and the noise in its estimation together determine sensor performance, and how device design and operating conditions can improve it.

The paper is organized as follows. Sections II–IV describe the devices and measurement protocol, present DC and AC field detection, and characterize the field-equivalent noise. Section V compares the measured performance with other magnetometer technologies and discusses routes toward improved sensitivity, followed by conclusions in Section VI. The Supplementary Material, available from the corresponding author upon request, provides calibration and signal-analysis details, additional amplitude-response measurements, and long-term stability analysis.

\section{Device and Measurement Method}

Figure~\ref{fig:setup}(a) shows the experimental setup for magnetic field sensing with a pMTJ. An FPGA and a custom daughter board generate electrical pulses and read the junction resistance state. An electromagnet beneath the pMTJ chip applies DC or AC magnetic fields perpendicular to the film plane. Each measurement cycle produces a binary outcome, and repeated outcomes are averaged to estimate a field-dependent state occupation probability.

Our samples are pMTJs with the same layer structure as those studied by Rehm \emph{et al.}~\cite{Rehm2021}, where the full stack is described. Each pMTJ has a composite CoFeB/W/CoFeB free layer (FL) separated by an MgO tunnel barrier from a CoFeB reference layer (RL), as illustrated schematically in Fig.~\ref{fig:setup}(b). The RL is ferromagnetically coupled to a synthetic antiferromagnet, which fixes its magnetization. The pMTJs have a nominal diameter of 50~nm and a free-layer thickness of 2~nm. Four pMTJs were studied: one for the DC magnetic field measurements and three, labeled MTJs J, K, and G, for the AC magnetic field measurements. When the magnetizations of the FL and RL are parallel (P), the junction is in a low-resistance state; when they are antiparallel (AP), the junction is in a high-resistance state. We take a positive applied field \(B\) to point along the RL magnetization, so that it favors the P state. We encode the measured magnetic state as a binary variable $b$, assigning $b=0$ to the AP state and $b=1$ to the P state. The P-state occupation probability, $p:=\Pr(b=1)$, is the probability that the junction is in the P state after the \textit{write} pulse.

The pMTJs were measured at room temperature using ground-signal-ground (GSG) probes connected to a custom analog daughter board. The setup was adapted from those described by Dubovskiy \emph{et al.}~\cite{Dubovskiy2024} and Criss \emph{et al.}~\cite{Criss2026}. The FPGA controlled the four-pulse sequence shown in Fig.~\ref{fig:setup}(b): \textit{reset}, \textit{verify}, \textit{write}, and \textit{read}. The \textit{reset} pulse initialized the junction in the P state, and a lower-amplitude \textit{verify} pulse confirmed successful initialization. The P and AP configurations correspond to energy minima separated by a barrier that suppresses spontaneous reversal on the timescale of a measurement cycle~\cite{Sun2026}. Following initialization, the FL magnetization samples a quasi-equilibrium Boltzmann distribution of orientations within the selected potential well.

The \textit{write} pulse then exerts a spin-transfer torque on the FL magnetization. Thermal fluctuations cause the initial orientation of the magnetization to vary from pulse to pulse and continue to influence its dynamics during the pulse~\cite{Rehm2023}. At a given pulse amplitude and duration, these fluctuations lead to a distribution of final states. The amplitude and duration of the \textit{write} pulse control the probabilities of obtaining the P and AP states. An applied magnetic field modifies the energy landscape, affecting both the initial distribution and the driven reversal dynamics, and thereby changes the switching probability. Following the \textit{write} pulse, the \textit{read} pulse determines the resulting state through the junction resistance. For the subsequent experiments, all four pulses had a duration of 10~ns. For measurements at a fixed operating point, the \textit{write} pulse amplitude was selected so that $p\approx0.5$ at zero coil current. The expected peak AC field was determined from the measured coil current using an independent gaussmeter calibration of the electromagnet (see the Supplementary Material).

The complete FPGA-controlled sequence repeats at 10.6~MHz and produces one binary outcome per measurement cycle, with the recorded outcomes then transmitted to a computer. For the AC measurements, we divided the binary data into non-overlapping bins of $N=10{,}656$ outcomes, each spanning approximately 1~ms. We estimated the P-state occupation probability in each bin by averaging its binary outcomes,
\begin{equation}
    p(t_j) = \frac{1}{N}\sum_{n=1}^{N} b_{j,n},
    \label{eqn:avg-ing}
\end{equation}
where $b_{j,n}\in\{0,1\}$ is the $n$th outcome in the $j^\text{th}$ bin and $t_j$ is the bin midpoint.

A static magnetic field shifts the mean state occupation probability at fixed \textit{write} pulse amplitude or changes the \textit{write} pulse amplitude required to obtain $p=0.5$. An AC field produces an oscillation in $p(t)$ at the applied frequency. These two responses form the basis of the DC and AC measurements described below.

\section{DC Magnetic Field Detection}

\begin{figure}[t]
  \centering

  \panel{0.88}{a}{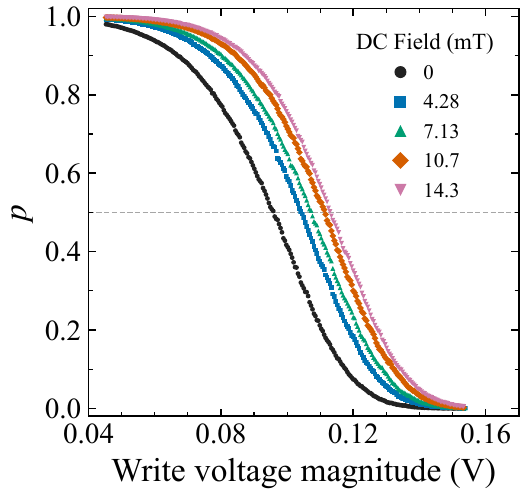}

  \par\smallskip

  \panel{0.88}{b}{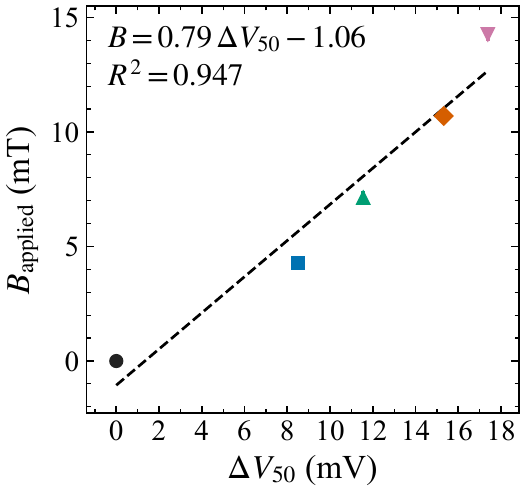}

  \caption{DC magnetic field response.
  (a) P-state occupation probability $p$ versus \textit{write} pulse voltage amplitude for several applied fields. (b) Applied field versus $\Delta V_{50}=V_{50}(B)-V_{50,\mathrm{ref}}$, with a free-intercept linear fit.}
  \label{fig:dc}
\end{figure}

To characterize the response to a DC magnetic field, we measured $p$ as a function of the \textit{write} pulse amplitude at several fixed applied fields. At each field, we swept the \textit{write} pulse amplitude from low to high over a range encompassing $p=0.5$. Figure~\ref{fig:dc}(a) shows the resulting curves for applied fields from 0 to 14.3~mT. Increasing the applied field shifts the curves toward larger \textit{write} pulse amplitude. This behavior is consistent with the applied field favoring the P state, so that a larger \textit{write} pulse amplitude is required to drive the junction into the AP state and recover the same probability $p$. This shift provides a means of measuring a static applied field using the stochastic response of the MTJ.

\begin{figure*}[t]
    \centering
        \centeredpanel{0.33}{a}{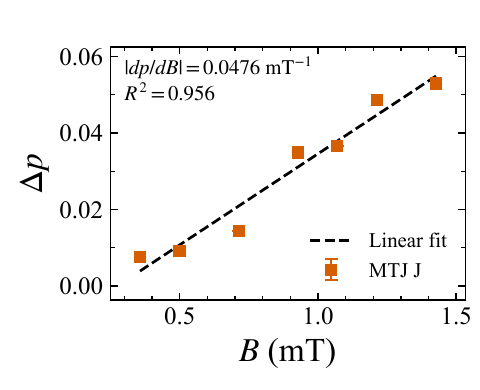}\hfill
        \centeredpanel{0.33}{b}{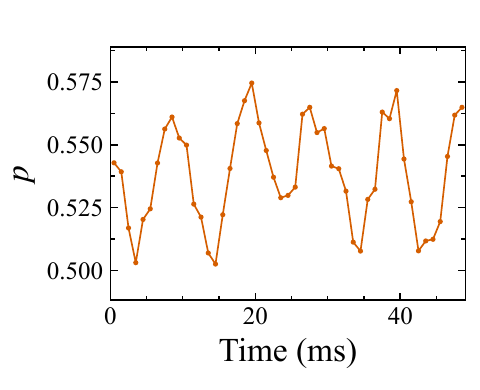}\hfill
        \centeredpanel{0.33}{c}{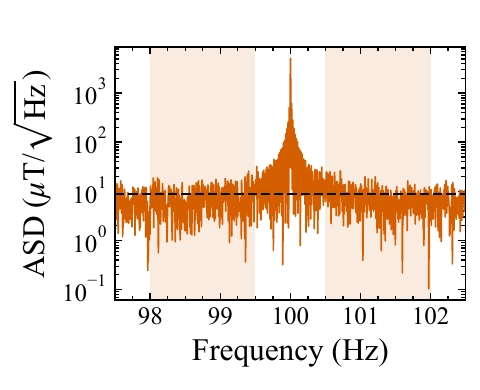}
    \caption{Calibration and representative AC detection. (a) Change in P-state occupation probability $\Delta p$ versus applied field $B$ for MTJ J at fixed \textit{write} pulse amplitude. Each probability change is measured relative to the average of the two adjacent zero-field measurements. Vertical error bars show $\pm 1$ standard error in $\Delta p$, propagated from the binomial counting uncertainties in the applied-field measurement and the two adjacent zero-field measurements. The error bars are smaller than the symbols. The slope of the linear fit gives $dp/dB$. (b) P-state occupation probability for MTJ J under a nominal 100~Hz field. The probability oscillates about a mean slightly above $0.5$, and the oscillation amplitude is converted to field units using the local slope $dp/dB$ from panel~(a). (c) Field-equivalent amplitude spectral density (ASD) for MTJ J near the 100~Hz drive frequency. The shaded bands mark the frequencies used to estimate the local noise. The dashed line gives the resulting noise ASD, approximately $9~\mu\mathrm{T}/\sqrt{\mathrm{Hz}}$, plotted for MTJ J at 100~Hz in Fig.~\ref{fig:frequency}(b).}
    \label{fig:ac_method}
\end{figure*}

We quantify the field-induced shift using $V_{50}$, the amplitude of the \textit{write} pulse voltage at which $p=0.5$. For each applied field, $V_{50}$ was extracted from a local linear fit over $0.3\leq p\leq0.7$. We then define the voltage shift relative to the zero applied field measurement as
\begin{equation}
    \Delta V_{50}(B) = V_{50}(B)-V_{50,\mathrm{ref}}.
\end{equation}
Here, $V_{50,\mathrm{ref}}$ is measured at zero applied field. This reference removes the baseline switching voltage without assuming that the total field at zero applied field vanishes.

Figure~\ref{fig:dc}(b) shows the applied magnetic field as a function of $\Delta V_{50}$. A free-intercept linear fit gives
\begin{equation}
    B = \left(0.79\,\frac{\mathrm{mT}}{\mathrm{mV}}\right) \Delta V_{50} - 1.06\,\mathrm{mT},
\end{equation}
with $R^2 = 0.947$. This relation converts the measured displacement of the switching curve into an estimate of the applied field. The fitted intercept of $-1.06\,\mathrm{mT}$ reflects the deviation of the data from a line through the origin, most visibly the large shift in $V_{50}$ between 0 and 4.28~mT. It should not be interpreted as a background field: because $\Delta V_{50}$ is measured relative to the zero-coil-current reference, the absolute background field cannot be inferred from this calibration. The rms residual of the fit, approximately 1.1~mT, indicates the field accuracy of this calibration over the measured range.

These measurements demonstrate a DC sensing method in which the applied field is inferred from a shift in the \textit{write} pulse amplitude required for $p=0.5$. The following section examines how an AC field can instead be detected through time-dependent changes in $p$ at a fixed \textit{write} pulse amplitude.

\section{AC Magnetic Field Detection}\label{sec:ac}

For AC detection, we held the \textit{write} pulse amplitude fixed near an operating point with $p\approx0.5$. The AC magnetic field produces an oscillation in the probability $p$. For sufficiently small field excursions, this response can be written as
\begin{equation}
    p(t) \simeq p_0 + \frac{dp}{dB} B_{\mathrm{peak}} \cos(2\pi ft+\phi),
\end{equation}
where $p_0$ ($\approx 0.5$) is the baseline P-state occupation probability, $dp/dB$ is the local occupation probability response to magnetic field, $f$ is the applied frequency, and $\phi$ is the phase.

\subsection{Calibration and Signal Extraction}
At the same \textit{write} pulse amplitude used for AC detection, we measured the change in P-state occupation probability $\Delta p$ as a function of applied static field $B$. For each field, we obtained $\Delta p$ by subtracting the mean of the two adjacent zero-field reference measurements from the measured probability. This procedure reduces the effect of baseline drift and is discussed in the Supplementary Material. Figure~\ref{fig:ac_method}(a) shows the resulting calibration for MTJ J. The slope of the linear fit gives the local field response $dp/dB$ at the operating point.

For an oscillating field, the peak field amplitude is
\begin{equation}
    B_{\mathrm{peak}} = \frac{\Delta p_{\mathrm{peak}}}{|dp/dB|},
    \label{eq:field_amplitude}
\end{equation}
where $\Delta p_{\mathrm{peak}}$ is the peak amplitude of the probability oscillation. This conversion assumes that the magnitude of the local field response remains equal to its DC value over the measured frequency range.

The signal frequency was estimated spectrally and refined from its phase drift relative to a reference sinusoid. We then performed quadrature demodulation in successive 30~s intervals, subtracting the mean probability within each interval. The cosine and sine quadratures, $X$ and $Y$, give the measured probability amplitude $\sqrt{X^2+Y^2}$. Details of the frequency estimation and demodulation are provided in the Supplementary Material.

Averaging the binary outcomes into probability bins of duration $\Delta t\approx1~\mathrm{ms}$ attenuates a sinusoidal signal. We corrected this attenuation using
\begin{equation}
    \Delta p_{\mathrm{peak}} = \frac{\sqrt{X^2+Y^2}} {|\operatorname{sinc}(f_{\mathrm{ref}}\Delta t)|}.
    \label{eq:bin_correction}
\end{equation}
Here, $f_{\mathrm{ref}}$ is the reference frequency used for demodulation, and $\operatorname{sinc}(x)=\sin(\pi x)/(\pi x)$. The corrected amplitude from each 30~s interval was converted to a magnetic-field amplitude using Eq.~\eqref{eq:field_amplitude}. These interval estimates were then averaged to obtain one field-amplitude estimate for each run.

Figure~\ref{fig:ac_method}(b) shows a representative probability trace for MTJ J under an applied 100~Hz field. The field-equivalent ASD of the full trace, shown in Fig.~\ref{fig:ac_method}(c), exhibits a peak at the drive frequency. We use the neighboring frequencies to estimate the local noise, as described below.

\subsection{Detection Across Frequency}

\begin{figure}[t]
  \centering

  \panel{0.90}{a}{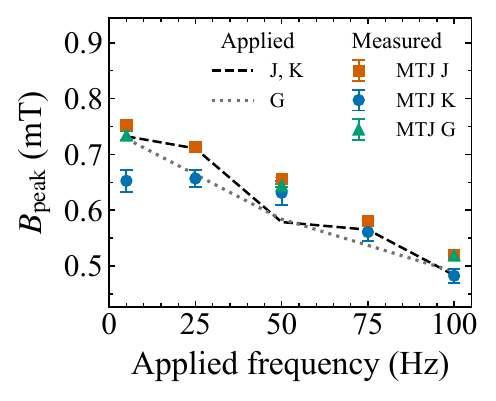}

  \par\medskip

  \panel{0.90}{b}{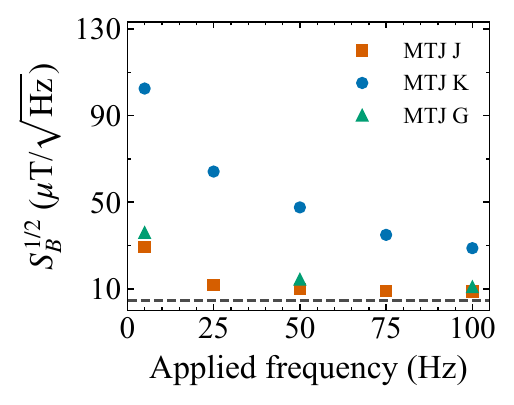}

  \caption{AC field detection using MTJs J, K, and G. (a) Inferred and expected peak field amplitudes versus applied frequency. Expected fields were determined from coil-current measurements and the independent electromagnet calibration. Error bars on the MTJ data show one sample standard deviation of the peak field amplitudes inferred from successive 30~s intervals within each run. (b) Local field-equivalent noise amplitude spectral density near each detection frequency. The horizontal dashed line marks the binomial benchmark for MTJ J, $4.6~\mu\mathrm{T}/\sqrt{\mathrm{Hz}}$. At 100~Hz, the measured noise ASD for MTJ J is approximately twice this benchmark.}
  \label{fig:frequency}
\end{figure}

We repeated the AC measurements at 5, 25, 50, 75, and 100~Hz on MTJs J and K, and at 5, 50, and 100~Hz on MTJ G. Figure~\ref{fig:frequency}(a) compares the inferred peak field amplitudes with the expected peak amplitudes, obtained as $\sqrt{2}$ times the rms field determined from the measured rms coil current and the electromagnet calibration. At fixed AC drive amplitude, the expected peak field decreased from approximately 0.73~mT at 5~Hz to 0.48~mT at 100~Hz. This decrease reflects the lower measured rms coil current at higher frequencies, consistent with the increasing inductive reactance of the coil.

The inferred amplitudes follow this decrease for all three devices. At 50~Hz, all three devices read 9--13\% above their respective expected amplitudes. The common excess at 50~Hz suggests that uncertainty in the reference field may contribute to the discrepancy. MTJ K reads approximately 11\% and 8\% below the expected amplitude at 5 and 25~Hz, respectively. All other measured amplitudes agree with their expected values to within 8\%.

These deviations are larger than the error bars, which reflect only the scatter between 30~s intervals within each run, suggesting a systematic contribution. For MTJ J, the relative standard error of the fitted field response $dp/dB$ is 9.6\%. We did not independently quantify the uncertainty in the rms coil-current measurement. The fitted coil calibration slope contributes a much smaller relative uncertainty of 0.35\% (see Supplementary Material). The expected fields also assume that the field per unit coil current is independent of frequency. AC measurements at two drive amplitudes for MTJ G at 5 and 100~Hz provide a further check of amplitude recovery and are presented in the Supplementary Material.

Figure~\ref{fig:frequency}(b) shows the local field-equivalent noise amplitude spectral density (ASD). The probability-noise ASD, $S_p^{1/2}(f)$, is converted to field units using
\begin{equation}
    S_B^{1/2}(f)
    = \frac{S_p^{1/2}(f)}{|dp/dB|}.
    \label{eq:asd}
\end{equation}
For each measurement, we calculated the one-sided probability power spectral density (PSD) of the mean-subtracted, Hann-windowed probability trace and corrected for attenuation due to probability binning. We averaged the corrected PSD over two frequency bands extending from 0.5 to 2~Hz below and above the measured signal frequency, excluding the driven peak. We then took the square root of this average and converted it to a field-equivalent noise ASD using Eq.~\eqref{eq:asd}.

For the 100~Hz measurement on MTJ J, the two averaging bands span approximately 98--99.5~Hz and 100.5--102~Hz, as shaded in Fig.~\ref{fig:ac_method}(c). This measurement yielded the lowest local field-equivalent noise ASD, approximately $9~\mu\mathrm{T}/\sqrt{\mathrm{Hz}}$. The dashed line in Fig.~\ref{fig:ac_method}(c) shows this value, which is plotted as the MTJ J point at 100~Hz in Fig.~\ref{fig:frequency}(b). Further details on the ASD calculation are provided in the Supplementary Material.

Extended acquisition times and amplitude stability are presented in the Supplementary Material. These measurements characterize the stability of the complete sensing measurement, including possible variations in the applied field and sensor response.

\section{Discussion}
These measurements demonstrate AC and DC magnetic field detection through changes in the probability of a pulse-actuated pMTJ being in the P or AP state. For DC detection, we sweep the \textit{write} pulse amplitude and measure how the applied field shifts the probability curve. For AC detection, we hold the \textit{write} pulse amplitude fixed and track oscillations in the P-state occupation probability caused by the alternating field.

\begin{figure}[t]
  \centering
  \begin{minipage}{1.00\linewidth}
    \plotfile{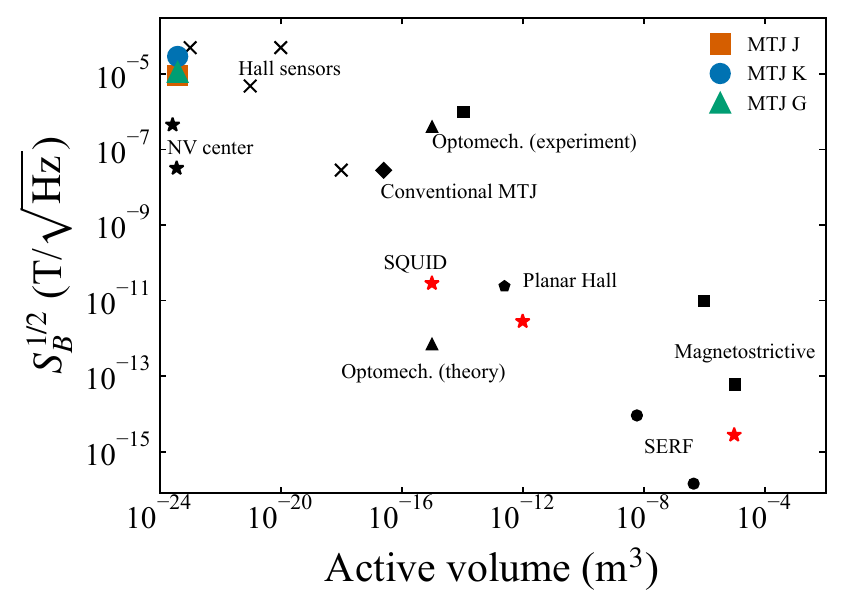}
  \end{minipage}
  \caption{Magnetic field sensitivity versus active volume for several magnetometer technologies. Sensitivity is expressed as field-equivalent noise ASD. For this work, volume denotes the magnetic free-layer volume; literature points use the reported sensor or detection volume. Volume definitions and operating conditions differ between technologies. Theoretical and experimental results are identified in the plot. Comparison adapted from Fig.~2 of Forstner et al.~\cite{Forstner2012}, with our pMTJ results and additional conventional MTJ and planar Hall sensor results added. Sources identified in that comparison are SERF magnetometers~\cite{Dang2010,Maser2011},
  SQUIDs~\cite{Kirtley1995,Faley2006,Baudenbacher2003},
  Hall sensors~\cite{Sandhu2004Micro,Sandhu2004Nano},
  NV centers~\cite{Maze2008,Balasubramanian2009},
  magnetostrictive sensors~\cite{Bucholtz1989,Dong2003,Osiander1996},
  and cavity optomechanical sensors~\cite{Forstner2012PRL,Forstner2012Model}. The conventional MTJ point uses the FFMTJ bridge of Davies et al.~\cite{Davies2021}, with noise ASD estimated from their Fig.~4 at 100~Hz and volume calculated as the summed free-layer volume of the bridge junctions. The planar Hall point uses the 50~nm thick sensor of Nhalil et al.~\cite{Nhalil2020}, with noise ASD reported at 50~Hz and volume calculated from the entire magnetic ellipse.}
  \label{fig:magnetometer_comparison}
\end{figure}

Figure~\ref{fig:magnetometer_comparison} compares the field-equivalent noise ASD of our pMTJs with representative magnetometer technologies as a function of active sensing volume. For our pMTJ, the active magnetic volume is the free-layer volume, approximately $3.9\times10^{-24}~\mathrm{m}^3$ for a diameter of 50~nm and thickness of 2~nm. Volume definitions and operating conditions differ between technologies, so the comparison indicates orders of magnitude rather than precise rankings.

Across technologies, the noise generally decreases as the sensing volume increases. The lowest noise levels, around $10^{-15}~\mathrm{T}/\sqrt{\mathrm{Hz}}$ and below, are reached by SERF magnetometers and SQUIDs with volumes of $10^{-8}~\mathrm{m}^3$ and larger. Sensors with volumes below $10^{-20}~\mathrm{m}^3$ have noise of approximately $10^{-8}~\mathrm{T}/\sqrt{\mathrm{Hz}}$ or higher. Our pMTJs lie at the small-volume end of this range, together with single NV centers. NV centers at comparable volume reach $3\times10^{-8}$ to $5\times10^{-7}~\mathrm{T}/\sqrt{\mathrm{Hz}}$, one to two and a half orders of magnitude below MTJ J, but they require optical initialization and readout and microwave control. The smallest Hall sensor in Fig.~\ref{fig:magnetometer_comparison} has a noise ASD of approximately $49~\mu\mathrm{T}/\sqrt{\mathrm{Hz}}$ at a volume of approximately $1.0\times10^{-23}~\mathrm{m}^3$, about 5.5 times the noise of MTJ J at about 2.6 times its volume. The conventional MTJ bridge of Davies \emph{et al.}~\cite{Davies2021} reaches lower noise, approximately $3\times10^{-8}~\mathrm{T}/\sqrt{\mathrm{Hz}}$, but with a summed free-layer volume about seven orders of magnitude larger.

The energy resolution per bandwidth, $E_R=S_B V/(2\mu_0)$, combines field noise and sensing volume in a single figure of merit~\cite{Mitchell2020}. Lines of constant $E_R$ have slope $-1/2$ in Fig.~\ref{fig:magnetometer_comparison}. For MTJ J at 100~Hz, $E_R\approx1\times10^{-28}~\mathrm{J\,s}$, or approximately $10^{6}\hbar$. By this measure, the pMTJ improves on the nanoscale Hall sensor ($\approx9\times10^{7}\hbar$) and the conventional MTJ bridge ($\approx7\times10^{7}\hbar$) and is comparable to the planar Hall sensor ($\approx5\times10^{5}\hbar$). The plotted NV centers and SERF magnetometers span approximately $10$--$2\times10^{3}\hbar$, and the SQUIDs span approximately $3\times10^{3}$--$3\times10^{5}\hbar$. The pMTJ thus has the smallest sensing volume among the electrically read sensors in Fig.~\ref{fig:magnetometer_comparison}, with an energy resolution better than other electrically read nanoscale sensors. If operation at 104~MHz reduced the measured noise ASD by the factor of 3.1 projected below for the binomial benchmark, $E_R$ would decrease by approximately a factor of ten.

For a quantitative noise benchmark, we assume statistically independent measurement cycles at a stable operating point and a linear field response that remains equal to its DC value over the measured frequency range. For an outcome rate \(r\) and mean P-state occupation probability \(p\), binomial statistics then give the one-sided field-equivalent noise ASD in the low-frequency limit as
\begin{equation}
S_{B,\mathrm{bin}}^{1/2}
=
\frac{\sqrt{2p(1-p)/r}}{|dp/dB|}.
\label{eq:binomial}
\end{equation}
Using \(r=10.6~\mathrm{MHz}\), \(p\simeq0.5\), and \(dp/dB=0.0476~\mathrm{mT}^{-1}\) for MTJ J gives approximately \(4.6~\mu\mathrm{T}/\sqrt{\mathrm{Hz}}\). Because the measured PSD is corrected for binning attenuation, we apply the same correction to this benchmark, which increases it by less than 2\% at 100~Hz. The measured minimum of approximately \(9~\mu\mathrm{T}/\sqrt{\mathrm{Hz}}\) for MTJ J at 100~Hz is therefore about twice this benchmark. This ratio does not depend on the calibrated value of \(dp/dB\), because the measured and benchmark noise both scale as \(1/|dp/dB|\).

Fluctuations in the operating probability $p$ on timescales longer than a measurement cycle can correlate successive outcomes and contribute to noise above the binomial benchmark. Possible sources include slow drift of the operating point, consistent with the rise of the noise toward low frequency in Fig.~\ref{fig:frequency}(b) and as observed in Ref.~\cite{Sidi2025}, noise in the \textit{write} pulse amplitude, magnetic field noise from the environment or the coil drive, and temperature fluctuations. If the DC calibration of Eq.~(3) also applies to MTJ J, write-pulse amplitude noise of approximately $10~\mu\mathrm{V}/\sqrt{\mathrm{Hz}}$ near 100~Hz would be sufficient to account for the observed excess.

Reductions in the field-equivalent noise are possible. At fixed $p$ and $dp/dB$, the binomial noise ASD scales as $r^{-1/2}$ for independent outcomes. Sidi El Valli \emph{et al.} demonstrated $104~\mathrm{MHz}$ operation with actuated pMTJs~\cite{Sidi2025}. Increasing $r$ from $10.6$ to $104~\mathrm{MHz}$ would lower the benchmark noise floor by a factor of $\sqrt{104/10.6}\simeq3.1$, to approximately $1.5~\mu\mathrm{T}/\sqrt{\mathrm{Hz}}$. Realizing this improvement would require the complete pulse sequence to fit within a 9.6~ns cycle, much shorter than the four 10~ns pulses used here. However, shorter \textit{write} pulses move operation toward the ballistic regime, where the model discussed below predicts a smaller $dp/dB$. A net reduction of the noise floor therefore requires $dp/dB$ to decrease by less than the factor of 3.1 gained from the higher outcome rate. This trade-off between outcome rate and field response remains to be studied experimentally.

Equation~\eqref{eq:asd} shows that field sensitivity can be improved by increasing the field response $|dp/dB|$ or by reducing probability noise. If the probability-noise ASD remains unchanged, doubling $|dp/dB|$ halves the field-equivalent noise ASD. For our actuated devices, the \textit{write} pulse amplitude and duration provide a way to tune this response. The relevant quantity to optimize is the field-equivalent noise, because changes in pulse settings can affect both the response and the probability fluctuations. For example, the binomial benchmark of Eq.~\eqref{eq:binomial} scales as $\sqrt{p(1-p)}/|dp/dB|$, so the optimal operating point need not be $p=0.5$. For the thermally activated form considered below, varying the pulse duration at fixed current and outcome rate minimizes this ratio near $p\approx0.2$, where it is approximately 14\% lower than at $p=0.5$.

For MTJ J, the calibration in Fig.~\ref{fig:ac_method}(a) gives $dp/dB\simeq0.048~\mathrm{mT}^{-1}$ near $p=0.5$. To examine how device design and pulse conditions affect this slope, we consider the uniaxial single-domain model of Liu \emph{et al.}~\cite{Liu2014}. The zero-field, zero-current energy barrier in units of the thermal energy $k_BT$ is
\begin{equation}
    \xi = \frac{K_{\mathrm{eff}}V}{k_BT} = \frac{\mu_0 M_s H_k V}{2k_BT},
\end{equation}
where $K_{\mathrm{eff}}$ is the effective perpendicular anisotropy energy density, $H_k$ is the corresponding anisotropy field, $M_s$ is the saturation magnetization, and $V$ is the free-layer volume. With the sign convention of Sec.~II, positive $B$ favors the initialized P state. For a \textit{write} pulse of current $I$ and duration $\tau$, the reduced drive is $i=I/I_{c0}-h$, where $h=B/(\mu_0H_k)$ is the reduced field and $I_{c0}$ is the zero-field, zero-temperature critical current. The field thus acts through $h$. Perpendicular anisotropy stabilizes the P and AP states equally, whereas the applied field favors one of them, so a smaller $H_k$ gives a larger $h$ for the same $B$. In the model, as in the experiment, the pulse amplitude is chosen to give $p=1/2$ at $B=0$ and is then held fixed as the field varies. We consider the two limiting regimes identified in Table~1 of Liu \emph{et al.}~\cite{Liu2014}.

In the subcritical, thermally activated limit ($i<1$), the effective barrier for escape from the P state is $\xi(1-i)^2$ [Eq.~(43) of Liu \emph{et al.}~\cite{Liu2014}]. At fixed current, a positive applied field raises this barrier. With effective attempt time $\tau_0^*$, the probability of remaining in the P state is
\begin{equation}
    p = \exp\!\left[-\frac{\tau}{\tau_0^*}e^{-\xi(1-i)^2}\right].
\end{equation}
Taking $\tau_0^*$ as locally field independent and differentiating at fixed $I$ and $\tau$ gives $dp/dB=p(-\ln p)\,2\xi(1-i)/(\mu_0H_k)$. At the midpoint, $-\ln p=\ln 2$, and the probability expression gives $\xi(1-i)^2=\ln[\tau/(\tau_0^*\ln 2)]$. Substitution yields
\begin{equation}
    \left.\frac{dp}{dB}\right|_{B=0,\,p=1/2} = \frac{\ln 2}{\mu_0 H_k}\sqrt{\xi\ln\!\left(\frac{\tau}{\tau_0^*\ln 2}\right)} .
    \label{eq:thermal_midpoint_slope}
\end{equation}
For $\tau=10$~ns and $\tau_0^*$ between 0.1 and 1~ns, the reduced barrier at the midpoint is only $\xi(1-i)^2\approx3$--$5$, so the Arrhenius form is only approximate. Taking $\xi=50$, the midpoint lies at $i\approx0.68$--$0.77$ over this range of attempt times.

At fixed $M_s$, $T$, $\tau$, and $\tau_0^*$, the slope scales as $\sqrt{\xi}/H_k\propto\sqrt{V/H_k}$. Lowering $H_k$ at fixed volume steepens the field response, whereas reducing the volume at fixed $H_k$ weakens it.

In the ballistic limit ($i\gg1$), spin torque drives reversal from a thermal distribution of initial magnetization angles. Let $P_{\mathrm{sw}}=1-p$ denote the switching probability and $\tau_D$ the intrinsic magnetic relaxation time. For $\xi\gg1$, combining the large-current trajectory, initial angular distribution, and final switching angle [Eqs.~(17), (20), and (72) of Liu \emph{et al.}~\cite{Liu2014}] gives
\begin{equation}
    -\ln P_{\mathrm{sw}} \simeq 4\xi\,\frac{(1+h)^2}{1-h}\exp\!\left[-2(i-1)\frac{\tau}{\tau_D}\right].
\end{equation}
At $B=0$, the logarithmic derivative of the rational factor is $3/(\mu_0H_k)$, and the field dependence of $i$ contributes $2\tau/(\mu_0H_k\tau_D)$ from the exponential. Using $dp/dB=P_{\mathrm{sw}}(-\ln P_{\mathrm{sw}})\,d\ln(-\ln P_{\mathrm{sw}})/dB$ at $P_{\mathrm{sw}}=1/2$, and eliminating $\tau/\tau_D$ with the midpoint condition $2(i-1)\tau/\tau_D=\ln(4\xi/\ln 2)$, gives
\begin{equation}
    \left.\frac{dp}{dB}\right|_{B=0,\,p=1/2} \simeq \frac{\ln 2}{2\mu_0 H_k}\left[3+\frac{\ln(4\xi/\ln 2)}{i-1}\right].
    \label{eq:ballistic_midpoint_slope}
\end{equation}
The first term dominates for $i-1\gg\ln(4\xi/\ln 2)/3$, which is approximately 2 for $\xi\approx50$. In this limit, the slope approaches $3\ln 2/(2\mu_0H_k)$, which has no explicit volume dependence. At fixed damping and pulse duration, $\tau_D\propto H_k^{-1}$ [Eq.~(9) of Liu \emph{et al.}~\cite{Liu2014}], so lowering $H_k$ increases $dp/dB$. Using the reported anisotropy field $\mu_0H_k\approx0.67$~T~\cite{Rehm2021} and taking $\xi=50$, Eq.~\eqref{eq:thermal_midpoint_slope} gives
$0.012$--$0.016~\mathrm{mT}^{-1}$, and the high-drive limit of Eq.~\eqref{eq:ballistic_midpoint_slope} gives $0.0016~\mathrm{mT}^{-1}$, compared with the measured $0.048~\mathrm{mT}^{-1}$ for MTJ J. The experimental value is thus larger than the predictions of the macrospin model.

Comparing the two macrospin model limits at the same $H_k$ shows that the switching regime itself strongly affects the slope. The ratio of Eq.~\eqref{eq:thermal_midpoint_slope} to the leading ballistic slope is $(2/3)\sqrt{\xi\ln[\tau/(\tau_0^*\ln 2)]}$, approximately 7.7--10.5 for $\xi=50$ and $\tau_0^*$ between 0.1 and 1~ns. Longer, lower-amplitude \textit{write} pulses that move operation toward the thermally activated regime therefore offer a steeper field response. Because the thermal slope depends on $\tau$ only logarithmically, the gain comes mainly from the change of regime. It must be weighed against the reduction in the outcome rate $r$, since the binomial benchmark scales as $1/(|dp/dB|\sqrt{r})$. The same trade-off applies to the higher outcome rates discussed above, because the shorter pulses they require move operation toward the ballistic regime. 

Both limits identify reduced perpendicular anisotropy as a route to a larger field response. Reducing the magnetic volume lowers the barrier as well, but it decreases the predicted thermal slope and leaves the leading ballistic slope unchanged. Lowering $H_k$ is limited by the requirement that the barrier still suppress spontaneous reversal over a measurement cycle. Beyond that point, the device becomes superparamagnetic, as discussed below.

A further reduction of the barrier can make spontaneous switching observable on the measurement timescale. This motivates the investigation of superparamagnetic MTJs, where spontaneous thermally driven transitions allow the field to be inferred from state probabilities and dwell times without repeated reset and \textit{write} pulses~\cite{Hirama2025,Liu2025}. Their performance depends on both the field response and the transition timescales: correlations between successive observations limit the number of statistically independent measurements, while the readout bandwidth limits how rapidly transitions can be resolved. The barrier and readout must therefore be considered together when optimizing sensitivity.

We also plan to investigate whether resonant activation~\cite{Doering1992}, in which barrier fluctuations influence escape times, can improve the field response of such devices. Any resulting enhancement must be assessed together with changes in probability noise to establish its benefit for magnetic field sensing.

\section{Conclusion}

We demonstrated DC and AC magnetic field detection using the state occupation probability of electrically actuated stochastic pMTJs. Static fields were inferred from shifts in the \textit{write} pulse amplitude required for $p=0.5$, while AC fields were recovered from probability oscillations at fixed pulse settings using digital lock-in analysis and a calibrated field response.

Measurements on three pMTJs demonstrated AC field detection from 5 to 100~Hz. The lowest field-equivalent noise amplitude spectral density, approximately $9~\mu\mathrm{T}/\sqrt{\mathrm{Hz}}$ at 100~Hz for MTJ J, is about twice the binomial benchmark for statistically independent measurement cycles at the 10.6~MHz measurement-cycle rate. This ratio does not depend on the field calibration. With a free-layer volume of approximately $3.9\times10^{-24}~\mathrm{m}^3$, the pMTJ has the smallest sensing volume among the electrically read sensors compared here.

Our analysis identifies several routes to improved sensitivity. The binomial benchmark decreases as $r^{-1/2}$ with the outcome rate, so operation at the 104~MHz rate already demonstrated with actuated pMTJs could lower it threefold, provided the field response is maintained. A macrospin model indicates that reduced perpendicular anisotropy and operation closer to the thermally activated switching regime both increase the field response, the latter at the cost of a lower outcome rate. Reducing the excess noise above the binomial benchmark, including the low-frequency noise observed in all three devices, would bring the measured noise closer to this limit.

\begin{acknowledgments}
We acknowledge support from the Office of Naval Research (ONR) under Award No. N00014-23-1-2771.
\end{acknowledgments}

\section*{Data Availability Statement}
The data and Supplementary Material that support the findings of this study are available upon reasonable request.

\makeatletter
\immediate\write\@auxout{\string\citation{MTJbibControl}}
\makeatother
\bibliographystyle{aipnum4-2}
\bibliography{sMTJ_Refs}
\end{document}